# Dynamics of Megaelectron Volt Electrons Observed in the Inner Belt by PROBA-V/EPT

**V. Pierrard[1,2], G. Lopez Rosson[1,2], and E. Botek[1]**

[1]Royal Belgian Institute for Space Aeronomy, Space Physics and STCE, Brussels, Belgium

[2]Université Catholique de Louvain, Center for Space Radiations, ELI-C, Louvain-La-Neuve, Belgium

Corresponding author: Viviane Pierrard Royal Belgian Institute for Space Aeronomy, Space Physics and STCE, 3 av. Circulaire, B-1180 Brussels, Belgium (viviane.pierrard@oma.be)

**Key Points**:

1. Five years of EPT observations show injections of energetic electrons in the slot region and in the inner belt during strong geomagnetic storms.
2. EPT detects MeV electrons present in the inner belt, especially in the South part of the South Atlantic Anomaly (SAA).
3. The penetration of the electrons in the SAA is strongly dependent on the geomagnetic activity level and on the energy of the electrons.

**Abstract**

Using the observations of the EPT (Energetic Particle Telescope) onboard the satellite PROBA-V, we study the dynamics of inner and outer belt electrons from 500 keV to 8 MeV during quiet periods and geomagnetic storms. This high time-resolution (2 sec) spectrometer operating at the altitude of 820 km on a low polar orbit is providing continuously valuable electrons fluxes for already 5 years. We emphasize especially that some MeV electrons are observed in low quantities in the inner belt, even during periods when they are not observed by Van Allen Probe (VAP). We show that they are not due to proton contamination but to clear injections of particles from the outer belt during strong geomagnetic storms of March and June 2015, and September 2017. Electrons with lower energy are injected also during less strong storms and the L-shell of the electron flux peak in the outer belt shifts inward with a high dependence on the electron energy. With the new high resolution EPT instrument, we can study the dynamics of relativistic electrons, including MeV electrons in the inner radiation belt, revealing how and when such electrons are injected into the inner belt and how long they reside there before being scattered into the Earth's atmosphere or lost by other mechanisms.

## 1. Introduction

Recently, NASA's Van Allen Probes (VAP) mission, launched in summer 2012 on an orbit close to the magnetic equatorial plane, showed that the expected population of MeV electrons in the inner belt was "missing" and that previous studies suggesting a long-lived, relatively static inner radiation belt of MeV electrons likely misidentified penetrating protons as inner belt electrons (Fennell et al., 2015). VAP observations of electrons made by the Energetic Particle, Composition, and Thermal Plasma/Magnetic Electron Ion Spectrometer (MagEIS) sensors showed no electrons > 900 keV with equatorial fluxes above background (i.e., > 0.1 electrons/(cm$^2$ s sr keV)) in the inner radiation zone (Fennell et al., 2015). More exactly, MagEIS detected some energetic electrons, but by recalibrating VAP electrons observations, the contamination due to protons in the inner belt was removed. Claudepierre et al. (2015) developed the algorithm to automatically remove background contamination from the Van Allen Probes MagEIS electron flux measurements. The major causes of contamination were found to be inner zone protons and bremsstrahlung X-rays generated by energetic electrons interacting with the spacecraft material.

This lack of MeV electron in the inner zone was a major result of VAP since previous empirical models like AE8 and AP9 (Vette, 1991) based on previous measurements in the radiation belts predicted the presence of significant fluxes of MeV electrons in the inner belt. This absence of MeV electrons in VAP observations during the first years of VAP observations could be explained by the contamination of protons in the inner zone in previous measurements, which could then make the models quite questionable in this region. But it could also be due to very low geomagnetic activity during solar cycle 23. Using MagEIS measurements on VAP from 0.7 to 1.5 MeV from April 2013 through September 2016, Claudepierre et al. (2017) show no electrons at all with an energy of 1.06 MeV in the inner zone before March 2015 and relatively few excursions of 1.06 MeV electrons into the inner slot region, where only the March and June 2015 events result in flux transport into the $L = 2.15$–$2.45$ range where L is the McIlwain (1966) parameter in Earth's radii. Li et al. (2015) used measurements from REPTILE (Relativistic Electron and Proton Telescope Integrated Little Experiment) on board Colorado Student Space Weather Experiment CubeSat in a low Earth orbit, to demonstrate that there exist sub-MeV electrons in the inner belt, but higher-energy electron (>1.6 MeV) measurements could not be distinguished from the background.

Using our EPT data, we can bring new answers to this question and provide new information on the dynamics of the energetic electrons in this region. We show that there are 1-2.4 MeV electrons in the inner zone, even before March 2015, and we can quantify these fluxes. Using EPT data from May 2013 up to 1 June 2018, we show MeV electron injections in the inner belt during March 2015 only up to L=1.4, corresponding to the South part of the South Atlantic Anomaly (SAA), and deeper penetration at lower L in June 2015 and also September 2017. These MeV electrons detected in the inner belt can clearly not be attributed to contamination from protons in the EPT instrument, since we can follow the flux injections during geomagnetic storms, as well as their lifetimes with our data. We show the energy dependence of the electrons in the dynamics of outer and inner belt, with the deepest injections for the strongest storms and for the lowest energies. The electrons in the inner belt remain trapped during several years with a flux slightly decreasing with time.

Section 2 briefly describes the instrument EPT and PROBA-V orbital characteristics and shows the flux variations with time observed by the instrument. In Section 3, we illustrate the flux maps observed by EPT at 820 km of altitude and discuss their time variations. In Section 4, we analyze the daily average flux evolution in the inner belt and discuss the results, before to conclude in Section 5.

## 2. Time variations in EPT Observations

The EPT (Energetic Particle Telescope) is a new detector especially designed to well discriminate the electrons, protons and Helium ions, so that it can make direct unambiguous high-quality measurements in the radiation belts and, specifically, in the inner zone despite the penetrating proton environment (Cyamukungu et al., 2014). This instrument was launched on the ESA satellite PROBA-V on 7 May 2013 to a LEO polar orbit (i=98.73°) at an altitude of 820 km, an orbit very different from VAP. This new EPT detector onboard PROBA-V provides high-resolution measurements of the charged particle radiation environment in space performing with direct electron (> 500 keV), proton (> 9.5 MeV) and helium ion (>38 MeV) discrimination (Pierrard et al., 2014). The time resolution is adjustable and nominally set to 2 seconds, which means that full spectra for each particle species is obtained every 2 s. The data are cleaned from possible pile-up in the detectors and suspicious data are removed from the database. Data filtering algorithms allow total noisy over-counting suppression and reducing of undercounting. More details concerning the validation and the filtering method applied to the data are described in the Technical Note of PROBA-V/EPT data production (Borisov and Cyamukungu, 2015) and in the article Cyamukungu et al. (2014). The observations of EPT are used to complete the TOP model (Benck et al., 2013) that gives the dynamics of the electron fluxes during quiet and geomagnetically disturbed periods at LEO orbit, also based on DEMETER and SAC-C satellite observations.

The instrument EPT has 6 different energy channels for electrons (see Pierrard et al., 2014). Especially useful for the present study, Channel 1 measures electrons from 500 to 600 keV, Channel 5 measures electrons from 1 to 2.4 MeV while Channel 6 measures the flux from 2.4 to 8 MeV.

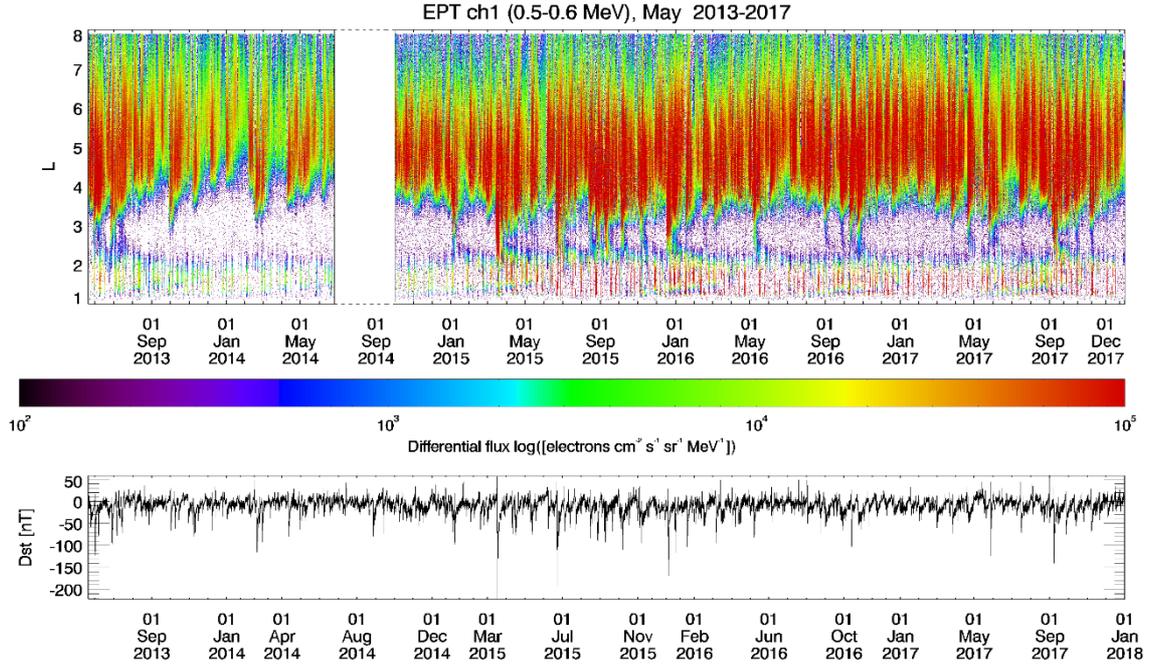

**Figure 1.** Electron flux measured by EPT in Channel 1 (500-600 keV) from 21 May 2013 up to 1 January 2018. The lower panel shows the Disturbed Storm Time (Dst) index during the same period.

Figure 1 illustrates our EPT observation of electron fluxes in Channel 1 (500-600 keV) from 21 May 2013 up to 1 January 2018. First, one can note that the fluxes of the electrons in the outer belt are very variable, especially during geomagnetic storms corresponding to inverted peak of Dst geomagnetic activity index. Such storms are characterized first by a dropout during the main phase, generally followed by a flux increase in the outer belt and injections in the slot region. Some injections even reach the inner belt during several storms, mainly during the more severe ones appearing in 2015 (Pierrard and Lopez Rosson, 2016), but also in 2016 and 2017. We can also note that an anomaly on sensor S3 noise of EPT was suddenly detected on 27 June 2014 and continued up to September 2014 (more specifically from 27/06/2014, 12:25:20 to 15/09/2014, 11:11:42). The EPT observations can't be used for scientific analyses during this period, represented on Fig. 1 by the white band. Some tests were immediately performed to check what was the origin of the problem and to recover the instrument nominal performance. The cause of the anomaly remains unknown: there was no SEP event or geomagnetic storm around those dates, and the possibility of electronics failure or voltage issues was discarded.

Before and after these dates, the EPT measurements give electron fluxes with a high resolution. The time period associated to the orbit of the spacecraft is 121 min. Some storm events can be noted in 2013 during the first months of the mission, but otherwise, 2013 and 2014 are very quiet with some small storms not penetrating to lower L than 3. The year 2015 is much more active, with major storms in January, March and June 2015, as studied in detail in Pierrard and Lopez Rosson (2016), but also during the second semester of 2015 where the slot region is regularly filled and electrons of 500-600 keV are injected in the inner belt. The fluxes decrease fast in the region between L=2.5 -3 and the new slot region typically come back after 10 days. But in the inner belt, the fluxes remain high and trapped during several months, with only slight decay. The loss of electrons in the slot region is expected to be due to VLF antennas (Lyons et al., 1972; Sauvaud et al., 2008). Note that even before the injections in 2015, the flux in the inner belt was not completely negligible in this energy range, even if clearly much lower than after the injections.

Figure 2 illustrates the electron flux measured from 21 May 2013 up to 1 January 2018 in Channel 5 (1-2.4 MeV) and in Channel 6 (2.4 - 8 MeV). While less frequent, injections also appear in Channel 5, increasing the flux in the inner belt (corresponding to the South Atlantic Anomaly (SAA) at these low altitudes). The flux in the inner belt increases only during the storms of March and June 2015 in this energy range, as well as during the storm of September 2017. The other storms also inject some MeV electrons at lower L in the slot region, but not as deep as in the inner belt at L<2 Re. Claudepierre et al. (2017) showed also with VAP data that some MeV electrons can be injected during strong geomagnetic storms.

Channels 2 to 4 (not illustrated here) show also the presence and the injection of energetic electrons in the inner belt during geomagnetic storms, especially during the strong St Patrick storm in March 2015 (Pierrard and Lopez Rosson, 2016).
On the contrary, no electron with an energy > 2.4 MeV is detected in the inner belt region corresponding to SAA at this low altitude (see panel 2 of Fig. 2). The flux of electrons with an energy > 2.4 MeV is increased highly in the horns during the St Patrick storm of March 2015, but not in the SAA, since the penetration for these ultra-relativistic electrons is limited to L= 2.6 Re during this event, which seems to be the deepest one of the 4-year period. This confirms the results of Baker et al. (2014), who showed with VAP that the ultra-relativistic electrons cannot penetrate in the inner belt. Cross calibration between the different detectors used for the different space missions can help to determine the limit in energy of electrons in the inner belt and provide many discoveries concerning the source and loss mechanisms of electrons (and protons) in the radiation belts.

The EPT results show that the injections in the inner belt are strongly energy dependent. Reeves et al. (2016) found that when MeV electrons are injected into the inner belt, they exist in a very narrow belt close to the Earth (L < 2 i.e. inside 2 Earth radii geocentric equatorial distance) while at 100 keV the belt is much wider, often extending to L > 3 and lasting much longer than their higher-energy counterparts. It seems that the flux is injected to deeper L when the storm is more active. This deepness depends on the energy, and is clearly deeper for lower energies, even inside the inner belt. Nevertheless, for electrons with E>2.4 MeV, no electrons are injected coming from the outer belt or slot region.

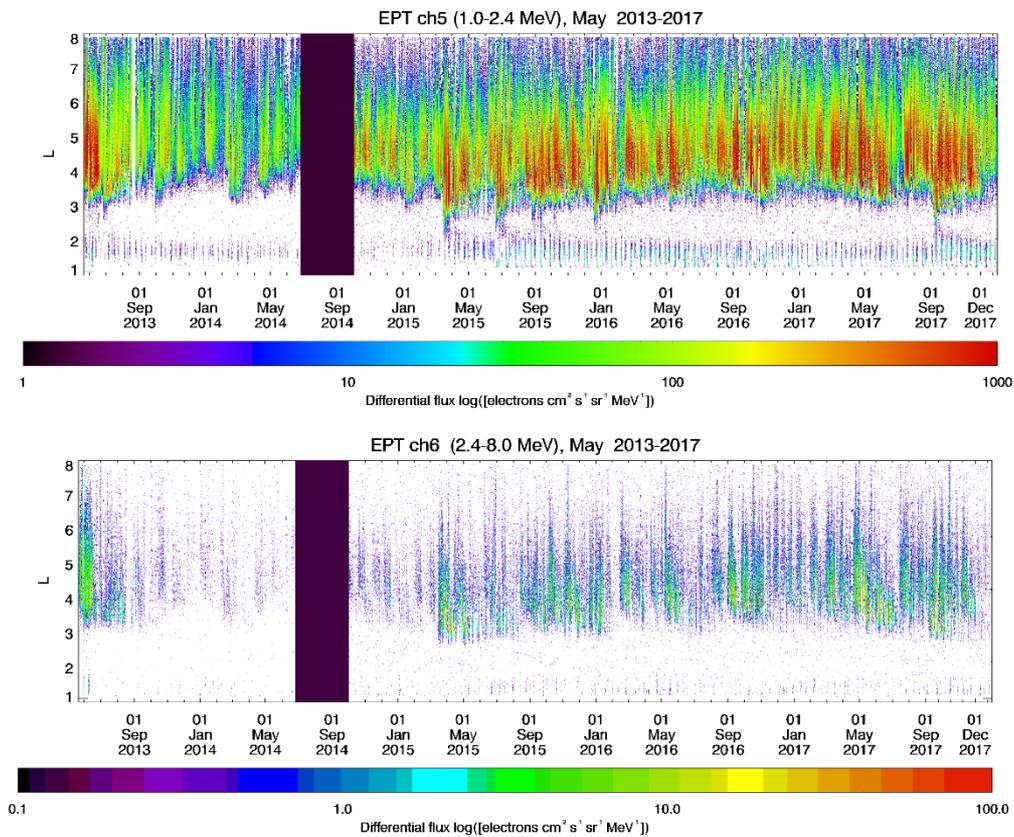

**Figure 2**. Electron flux measured by EPT in Channel 5 (1-2.4 MeV) (upper panel) and Channel 6 (2.4-8 MeV) (bottom panel) from 21 May 2013 up to 1 January 2018.

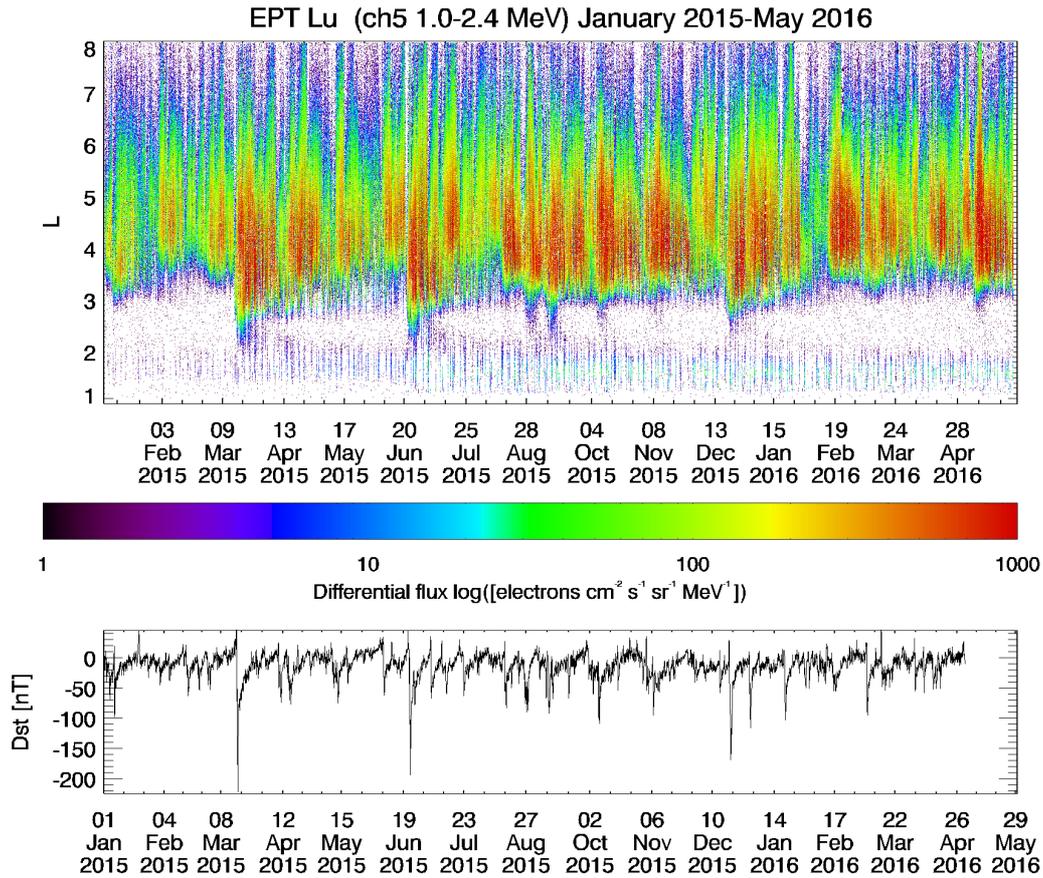

**Figure 3.** Electron flux measured by EPT in Channel 5 (1-2.4 MeV) from 1st Jan 2015 up to May 2016. The bottom panel illustrates the Dst index during the same period.

Figure 3 shows electron flux measured by EPT in Channel 5 (1-2.4 MeV) during the active year 2015 and up to May 2016. The bottom panel illustrates the Dst (Disturbance storm time) index during the same period and shows the clear correspondence between the injections and the strongest geomagnetic storms characterized by Dst indices < -100 nT. We can note also on this figure that after the injection of June 2015, the MeV electrons of the inner belt reached L values as low as L=1.1 while no MeV electrons were observed at L< 1.4 before this event.

## 3. Electron flux maps

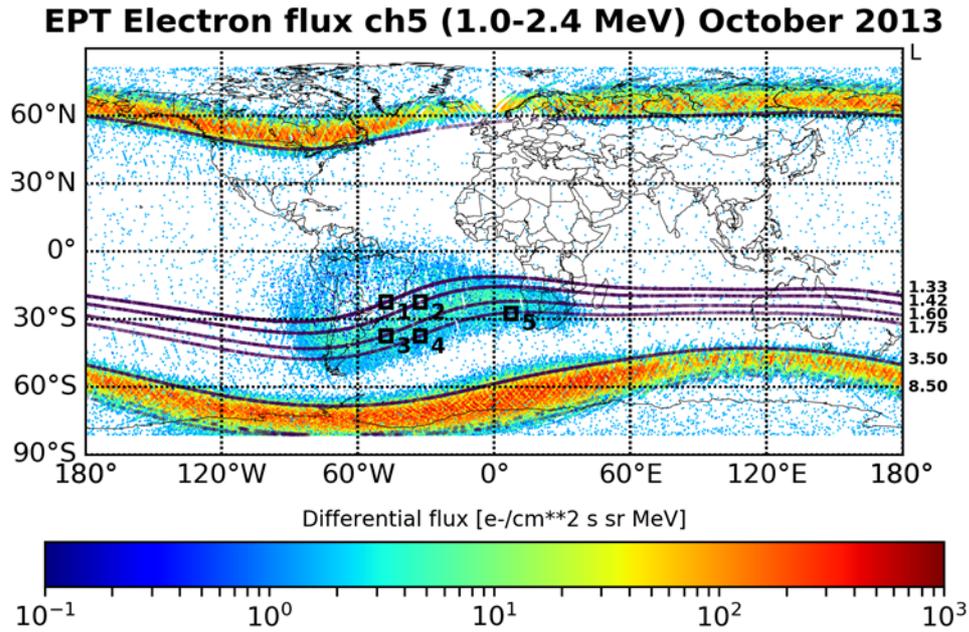

**Figure 4**. Electron fluxes observed by EPT in channel 5 (1.0-2.4 MeV) during October 2013. The 5 bins used to determine electron flux variations are also illustrated, as well as constant L (black) lines corresponding to L=1.33 (closer to equator), 1.42, 1.6, 1.75 (inner belt in the South hemisphere), L= 3.5 and L=8.5 (higher latitudes corresponding to the outer belt).

Figure 4 illustrates the map of electron fluxes observed during the month of October 2013 in Channel 5 (1-2.4 MeV). The inner belt is clearly visible as the South Atlantic Anomaly (SAA). The fluxes are slightly higher in the South part of the SAA, the limit between the South part (with higher fluxes in the SAA) and the North part of the SAA being located at L=1.4. Like for other energies, the fluxes are nevertheless much higher in the high latitude horns corresponding to the penetration of the outer belt at low altitudes. Between the inner and the outer belt, the slot region with lower fluxes appears typically between L=2 and L=3 in this energy range. The fluxes are almost constant along constant L values in the outer belt, but very different along L in the inner belt depending if the satellite at the low altitude of 820 km is located inside or outside the SAA. Using a lower minimum threshold for the color scale, we can observe the background and the lack of electrons in the counterpart of the SAA in the North hemisphere (Pierrard et al., 2014). Note nevertheless that the gap in the Northern horn close to Europe is due to the data transmission to the Redu ground station.

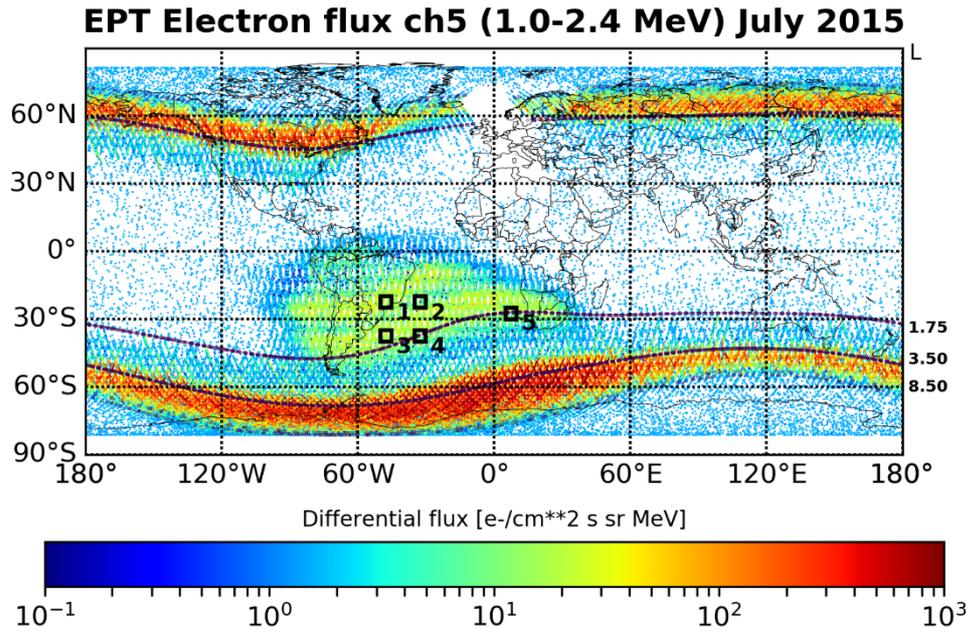

**Figure 5**. Electron fluxes observed by EPT in Channel 5 (1-2.4 MeV) during July 2015.

Figure 5 illustrates the month of July 2015 (after the storm of June 22-23) for electrons between 1 and 2.4 MeV. The slot region is filled, as well as all the regions of the SAA. One can see that electrons with E>1 MeV reach L<1.4 so that higher fluxes appear also in the North part of SAA, even if lower fluxes remain observable along L=1.4 and L<1.22. These high fluxes are still visible a few weeks after this event, and also after the event of September 2017, while they are not observable during other time periods. In March 2015 for instance, the flux has increased and MeV electrons are injected in the South part of the SAA and in the slot region, but not in the North part of SAA. This is clearly visible also in Fig. 3 where it can be seen that the fluxes at L<1.4 are not modified during the storm of March 2015, but increases during the event of June 2015.

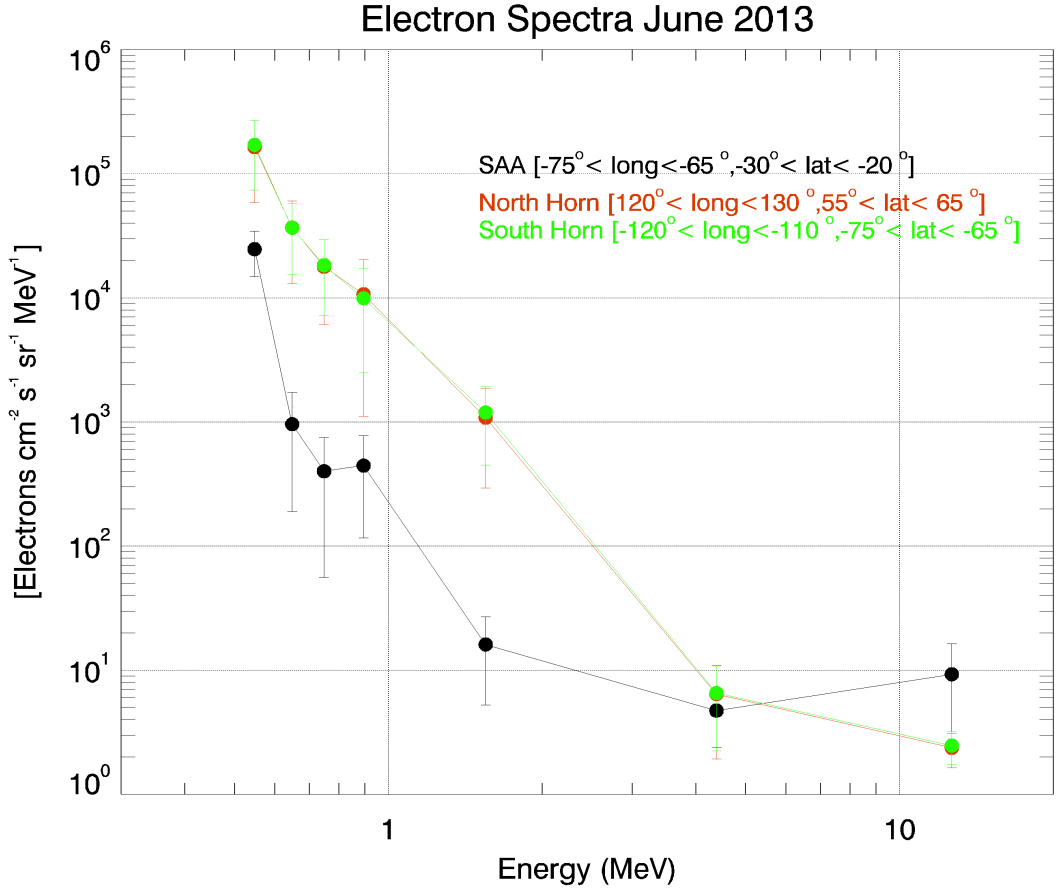

**Figure 6.** The averaged differential spectra observed by EPT in June 2013 in the North horn (red), South horn (green) and in the SAA (black).

The fluxes are thus very variable, especially in the outer belt but also in the inner belt corresponding to the SAA in our case of a LEO satellite. But the electron fluxes are always much higher for all energies in the horns than in the SAA. In Fig. 6, we illustrate the average differential spectra obtained with the EPT measurements during the month of June 2013 (low activity) in the horns and in the SAA using the different following bins:

For the horns, a square of 10°x10° was used.

In the North horn (in red) at [120°<Long < 130°], [55°< Lat < 65°], and in the South horn (in green) at [-120°< Long < -110°], [-75°< Lat < -65°].

Considering that the fluxes are much lower in the SAA, a same size bin is used, located at [-75°< Long < -65°], [-30°<Lat < -20°]. In the horns from the North and South hemisphere, the flux differences are not very high. Error bars corresponding to the standard deviation are also illustrated.

Note that in June 2015, the fluxes have increased in the SAA, especially in Channels 1 to 4, reaching then similar values in the horns and in the SAA for the energy range between [500-600 keV]. But the fluxes observed in the SAA are very low, especially for > 1 MeV electrons.

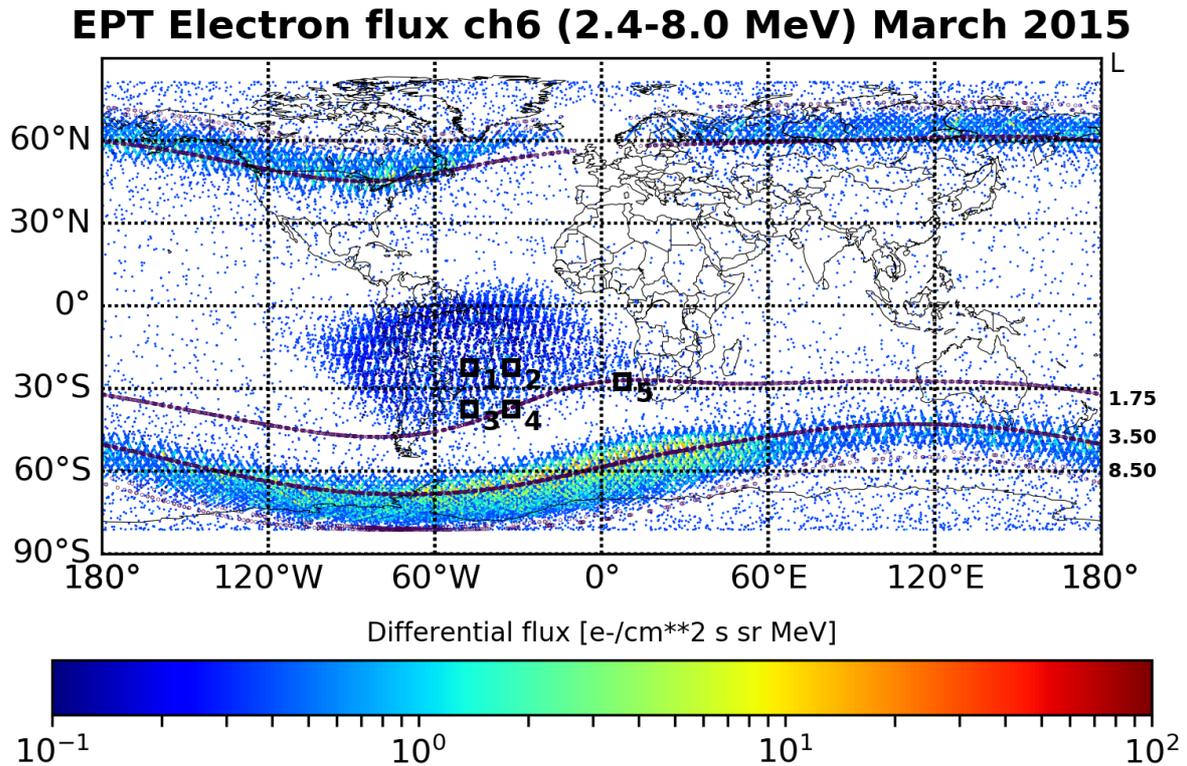

Figure 7. Electron fluxes observed by EPT in Channel 6 (2.4-8 MeV) during March 2015.

Figure 7 illustrates the electron fluxes observed by EPT in channel 6 (2.4-8 MeV) during the month of March 2015. The flux observed by EPT in this energy range is almost negligible in the SAA, but its sketch is visible when accumulating on one month since the measured flux is not exactly zero (a dark blue dot appears each time a flux measurement is not null). The high latitude fluxes are higher than in the SAA (Katsiyannis et al., 2018).

**4. Daily average flux in the SAA**

To study the time evolution of the electron population in the SAA (and thus in the inner belt), we proceed by determining 5 bins of 5° square. The five bins are illustrated in Fig. 4 and 5, as well as some constant L lines.

Their positions correspond to:

Bin 1: [-50 °< Long < -45°], [-25° < Lat < -20°]

Bin 2: [-35° < Long < -30°], [-25° < Lat < -20°]

Bin 3: [-50° < Long < -45°], [-40° < Lat < -35°]

Bin 4: [-35° < Long < -30°], [-40° < Lat < -35°]

Bin 5: [ 5° < Long < 10°], [-30° < Lat < -25°]

The centers of the bins correspond to:

Bin 1: L ∼ 1.33 ; B ∼ 0.166 nT

Bin 2: L ∼ 1.42 ; B ∼ 0.174 nT

Bin 3: L ∼ 1.60 ; B ∼ 0.178 nT

Bin 4: L ∼ 1.75 ; B ∼ 0.181 nT

Bin 5: L ∼ 1.75 ; B ∼ 0.197 nT

In these bins, we determine daily averaged flux observed by EPT. This is illustrated in Fig. 8 for Ch 1 (500-600 keV in black), Ch 5 (1-2.4 MeV in red) and Ch 6 (2.4-8 MeV) in green from 21 May 2013 up to June 2018. Top panel corresponds to Bin 1, middle panel to Bin 4 and Bottom panel to Bin 5. Bins 2 and 3 are not shown here since they give very similar results as Bin 1 and 4 respectively. The blue vertical lines correspond to SEP events.

The fluxes of higher energy are of course lower than the flux in Channel 1, but both Channels 1 and 5 follow the same flux increase starting in January 2015. This increase seems to be associated to the first geomagnetic storms visible on that time in January 2015, clearly not to SEP contrary to what happened for protons (Lopez Rosson and Pierrard, 2017). The increase is regular in Bins 4 and 5 located in the South part of SAA, even if the small sudden increases are visible during some important storms. The flux in Bins 4 and 5 (Ch 1 and 5) slightly decreases after June 2015, but a sharp increase is again visible after the big storm of 7 September 2017, resulting again in a slow decrease of the fluxes during the next months and in 2018.

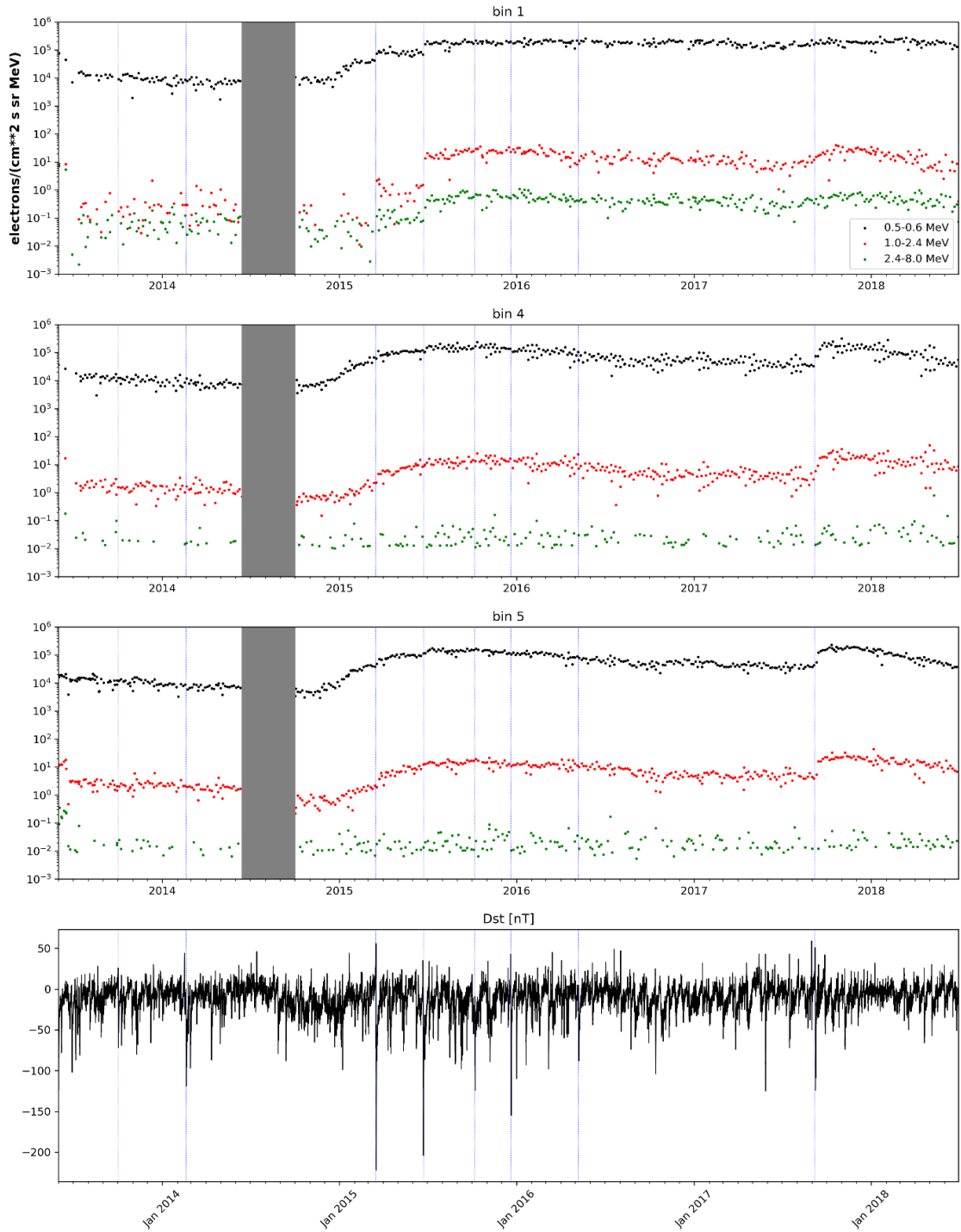

**Figure 8**. Daily-averaged electron flux observed by EPT every day for Ch 1 (500-600 keV in black), Ch 5 (1-2.4 MeV in red) and Ch 6 (2.4-8 MeV in green) from 1 June 2013 up to 1 June 2018 in the different bins. Top panel corresponds to Bin 1, middle panel to Bin 4 and bottom panel to Bin 5. Storms of Jan, March and June 2015, as well as Sep 2017 lead to an increase of the flux (for Ch 1 and Ch 5). The electrons remain trapped during several months since the flux decreases very slowly in all bins.

In Bin 1 located in the North part of the SAA where the fluxes are generally very low for ultra-relativistic electrons, the flux increase of June 2015 is especially visible for > 1 MeV electrons. It seems that the energetic electron fluxes in these regions sharply increased during the very strong events of March and June 2015, contrary to what is observed in Bins 3, 4 and 5 where the increase is more gradual and where no change is observed for E>2.4 MeV. The effect of the storm of September 2017 is less visible for the very high energies, but very clear for lower energy ranges in all bins.

## 5. Discussion and conclusions

New high resolution PROBA-V/EPT measurements of the electrons fluxes allow us to better understand the dynamics of the electrons in the inner and outer radiation belts. EPT put in perspective the VAP discovery the expected population of MeV electrons in the inner belt was missing and that previous studies suggesting a long-lived, relatively static inner radiation belts likely misidentified penetrating protons as inner belt electrons (Fennell et al., 2015). EPT shows indeed extremely low MeV fluxes before 2015. Later, electron MeV fluxes slightly increased after some injections in 2015 and 2017. The threshold of the impenetrable barrier discovered by Baker et al. (2014) is thus a little bit higher than 1 MeV since some injections are observed at this energy, but not at higher energies than 2.4 MeV. The injection observations of MeV electrons by EPT cannot be due to proton contamination since injections associated to geomagnetic storms are clearly identified. The new capability of EPT to well discriminate the particle species and energy ranges led to new discoveries concerning their source and loss mechanisms regarding the dynamics in this region.

MeV electrons exist in the inner belt, but electrons with E>2.8 MeV are very rare and not due to injections by geomagnetic storms. EPT reveals the dynamics of MeV electrons in the inner radiation belt showing how and when MeV electrons are injected into the inner belt and how long they reside there before being lost. The observed solar cycle variation observed in the radiation belts (Miyoshi et al., 2004) is clearly due to the higher number of geomagnetic storms during solar maximum injecting electrons in the inner and outer belts.

Electrons can be transported Earthward either through the action of large-scale electric fields (convection) or through the action of stochastic/diffusive fluctuations in the fields. As electrons are transported Earthward, they also gain energy through betatron and Fermi acceleration in the Earth's magnetic field. As is apparent from Fig. 1 and 2, whatever processes control the transport (and energization), they are energy-dependent and vary from event to event with some events producing deeper penetration and higher-energy penetration than others.

The residence lifetime of electrons in the radiation belts is also strongly energy- and activity-dependent (e.g. Ripoll et al., 2016). Electrons can be lost from the radiation belts either through outward radial transport or by scattering the electrons along the magnetic field lines into the atmosphere. Scattering of particles in the atmosphere occurring in between the belts is caused by resonant interaction with electromagnetic waves which in turn are produced by instabilities in thermal plasmas. With its high resolution measurements, EPT can help to determine the dynamics of the electron fluxes in the outer and the inner belts, and especially for MeV electrons that are injected only rarely (but not never) at L<3.

**Acknowledgements**

E. Botek thanks funding from the Research Executive Agency under the European Union's Horizon 2020 Research and Innovation program (project ESC2RAD: Enabling Smart Computations to study space RADiation effects, Grant Agreement *776410*).

The authors thank S. Benck, S. Borisov and M. Cyamukungu for the validation of the EPT data at CSR/UCL. The EPT data are available on Space Situational Awareness website of ESA: http://swe.ssa.esa.int/space-radiation (EPT).